\documentstyle[a4,12pt,epsf]{article}

\newcommand{\noi}{\noindent}
\newcommand{\eq}{\begin{equation}}
\newcommand{\en}{\end{equation}}
\newcommand{\eqa}{\begin{eqnarray}}
\newcommand{\ena}{\end{eqnarray}}

\begin{document}

\renewcommand{\theequation}{\arabic{section}.\arabic{equation}}
\renewcommand{\thesection}{\arabic{section}}

\hbox{}
\noindent December 1999  \hfill JINR E2--99-288

                            \hfill HUB--EP--99/51

\vspace*{1.0cm}

\begin{center}

{\Large
Fermionic correlators and zero--momentum modes in quenched lattice QED}
\vspace*{0.8cm}

I.L. Bogolubsky$^a$, V.K.~Mitrjushkin$^a$,
M. M\"uller--Preussker$^b$, P. Peter$^b$ \\
and N. V. Zverev$^b$

\vspace*{0.3cm}

\end{center}

{\sl \noindent
 \hspace*{6mm} $^a$ Joint Institute for Nuclear Research, Dubna, Russia \\
 \hspace*{6mm} $^b$ Institut f\"ur Physik, Humboldt-Universit\"at zu Berlin,
                   Germany}

\vspace{0.5cm}
\begin{center}

\vspace{1cm}
{\bf Abstract}
\end{center}

For the Lorentz gauge the influence of various Gribov gauge copies on the
fermion propagator is investigated in quenched compact
lattice QED. Within the Coulomb phase besides double Dirac sheets
the zero-momentum modes of the gauge fields are shown to cause
the propagator to deviate strongly from the perturbatively expected
behaviour. The standard way to extract the fermion mass fails.
The recently proposed zero-momentum Lorentz gauge is demonstrated
to cure the problem.

\section{Introduction}
\setcounter{equation}{0}

Lattice gauge theories allow to compute most of the
relevant observables without any gauge fixing. Nevertheless, computations of
gauge dependent objects, e.g. gauge or fermion correlators, can
give us more detailed information about nonperturbative
properties of quantum fields and allow a direct comparison with
the perturbation theory in the continuum.

However, it is well known that gauge fixing, in particular the Lorentz (or
Landau) gauge, leads to the occurence of gauge or Gribov copies \cite{Grib}.
For QED this happens even in the continuum, as long as the theory is
defined with toroidal boundary conditions \cite{Killingb}.

In this paper we are going to consider compact $U(1)$ lattice gauge
theory with Wilson fermions \cite{Wilson} in the quenched 
approximation (qQED). 
In the weak coupling region, i.e. within the Coulomb phase,
it describes massless photons (weakly) interacting with fermions.
Full lattice QED is expected to reproduce continuum perturbative QED,
which has been experimentally proven with greatest precision.

The standard iterative way to fix the Lorentz gauge for compact $U(1)$
lattice gauge theory has been shown to lead to serious Gribov copy effects
\cite{NaPl,NaSi,BoMiMPPa,PhcrHetr}. As a consequence
the transverse non-zero momentum photon correlator does not reproduce
the perturbatively expected zero-mass behaviour.
For the fermion correlator a strong dependence on the achieved gauge copies
has been reported, too \cite{NaSi}. The standard fermion mass determination
becomes badly defined. Careful numerical \cite{BoMiMPPa,PhcrHetr,BoMiMPPe,
BoMiDD} and analytical \cite{Mitr1,Mitr2} studies have shown that the main
gauge field excitations responsible for the occurence of disturbing
gauge copies are {\it double Dirac sheets} (DDS) and {\it zero-momentum modes}
(ZMM).

For the aim of achieving a better agreement between numerical lattice results
and lattice perturbation theory, the Lorentz gauge fixing procedure was
coupled with additional gauge fixing conditions.
The authors of \cite{NaSi} proposed to employ a unique initial gauge
realized by the "maximal tree" axial gauge condition. In \cite{BoMiMPPa} a
non-periodic gauge determined from spatial Polyakov loop averages was
proposed to suppress unwanted DDS. It turned out that a gauge which
allows to suppress DDS is sufficient to produce a correct non-zero
momentum photon correlator perfectly compatible with a vanishing
photon mass. However, in this way the Gribov problem is not yet solved,
because the removal of DDS does not automatically lead to the global
extremum of the Lorentz gauge functional. As we have seen recently
\cite{BoMiMPPe}, the above mentioned axial gauge even fails to remove DDS.
In the same paper we have shown that in order to solve the
Gribov problem the ZMM of the gauge fields have to
be necessarily suppressed, too. An alternating combination
of the Lorentz gauge fixing steps with non-periodic gauge transformations
suppressing ZMM provides a practical solution of the problem.

In the present paper, we are mainly studying the influence of the ZMM
on the fermion correlator when applying the Lorentz gauge.
The aim is to check whether the
results of \cite{BoMiMPPa,PhcrHetr,BoMiMPPe,BoMiDD,Mitr1,Mitr2}
apply to the fermion correlator, too. DDS will be removed from
the beginning in order to concentrate on
the effect of the ZMM. We shall demonstrate that only the correct account
of the ZMM will allow us to determine the fermion mass properly.

The outline of our paper is as follows. In Section 2 we introduce
our notations in particular for the fermion propagator. For the latter
we provide the analytical representation in a fixed constant gauge
field background. Section 3 presents the details of the gauge fixing
procedures employed. In Section 4 Monte Carlo results for the fermion
correlator obtained with different Lorentz gauge fixing procedures are
compared with the zero-momentum mode approximation of the correlator, for
which only the constant background gauge field modes are taken into account.
The conclusions are drawn in Section 5.

\section{The Fermion Correlator}
\setcounter{equation}{0}

We consider 4d compact U(1) gauge theory on a finite lattice
with size $V=N_s^3 \times N_t$.
The standard Wilson action consists of the pure gauge $S_G$ and
the fermion contribution $S_F$ as follows
\eq \label{1}
S_G = \beta\sum_{x,\mu < \nu}\left(1-\cos\theta_{x,\mu\nu}\right).
\en
$\theta_{x,\mu\nu}=\theta_{x,\mu}+\theta_{x+\hat{\mu},
\nu}-\theta_{x+\hat{\nu},\mu}-\theta_{x,\nu}$ is the plaquette angle with
$\theta_{x,\mu}\in (-\pi, \pi]$ denoting the link gauge field variable,
$\beta=1/e_0^2$ is the inverse coupling.
The lattice spacing is put $a=1$.
\\
The fermion part is given by
\eqa \label{2}
S_F&=&\sum_{x,y}\overline{\psi}_{x} {\bf M}_{xy}(\theta) \psi_{y} \\
   &=&(4+m_0)\sum_{x}\overline{\psi}_{x}\psi_{x}
    - \frac{1}{2}
\sum_{x,\mu}\Bigl\{\overline{\psi}_{x}{\rm e}^{{\rm i}\theta_{x,\mu}}
(1-\gamma_\mu)\psi_{x+\hat{\mu}} + \overline{\psi}_{x+\hat{\mu}}
{\rm e}^{-{\rm i}\theta_{x,\mu}}(1+\gamma_\mu)\psi_{x}\Bigr\},
\nonumber
\ena
$\psi_x$ denoting the fermion field, $\gamma_\mu$ the Hermitian
Dirac matrices.

For the gauge and the fermion field we apply periodic boundary conditions
(b.c.) except for the fermion field in the imaginary time direction
$~x_4$, where we prefer to use anti-periodic ones.

We are going to study the fermion correlator for given gauge fields
$\theta_{x,\mu}$
\eq \label{4a}
\Gamma(\tau;\theta)=\frac{1}{V}\sum_{\vec{x},x_4}\sum_{\vec{y}}
{\bf M}^{-1}_{\vec{x},x_4; \vec{y},y_4}(\theta), \qquad y_4 = x_4 + \tau.
\en
For simplicity, we restrict ourselves to the scalar and vector parts
  of the fermion correlator, respectively

\eqa \label{4b}
\nonumber
\Gamma_S(\tau;\theta) &=&\frac{1}{4}{\rm Re\,Tr\,}
\left(\Gamma(\tau;\theta)\right);
\\
\Gamma_V(\tau;\theta) &=&\frac{1}{4}{\rm Re\,Tr\,}
\left(\gamma_4\Gamma(\tau;\theta)\right),
\ena
where the trace is taken with respect to the spinor indices.
For anti-periodic b.c. in $~x_4~$ the vector (scalar) part
becomes an even (odd) function in $~\tau~$ around $~\tau = N_t / 2~$, for
periodic b.c. vice versa.

In qQED the above correlator has to be averaged with respect
to the gauge field $~\theta_{x,\mu}~$ with the weight $~\exp(-S_G)~$.
Lateron, we shall compare the quantum average
$~\langle~\Gamma~\rangle_{\theta}~$ with
the zero-momentum mode approximation
where only background gauge fields being constant
in space-time are taken into account. Therefore, we construct analytically
the correlator for a (uniform) gauge configuration given by
\eq \label{5}
\theta_{x,\mu}\equiv\phi_{\mu}, \qquad  -\pi < \phi_{\mu} \le \pi,
            ~\mu=1,\cdots,4 .
\en

One obtains the following finite size results for the
scalar and vector parts, respectively

\eqa \label{6a}
\Gamma_S(\tau;\phi) &=& \frac{\delta_{\tau,0}}{2(1+{\cal M})} 
- \frac{1+{\cal E}^2-2{\cal E}(1+
{\cal M})}{1-{\cal E}^2} \times \\
&\times& \frac{[{\cal E}^{\tau}-{\cal E}^{2N_t-\tau}]\cos(\phi_4\tau)-
c[{\cal E}^{N_t+\tau}-{\cal E}^{N_t-\tau}]\cos[\phi_4(N_t-\tau)]}
{1+{\cal E}^{2N_t}-2c{\cal E}^{N_t}\cos(\phi_4 N_t)},
\nonumber
\ena

\eqa  \label{6b}
\Gamma_V(\tau;\phi) &=& \frac{1-\delta_{\tau,0}} {2(1+{\cal M})}
\times \\
&\times& \frac{[{\cal E}^{\tau}+{\cal E}^{2N_t-\tau}]\cos(\phi_4\tau)
-c[{\cal E}^{N_t+\tau}+{\cal E}^{N_t-\tau}]\cos[\phi_4(N_t-\tau)]}
{1+{\cal E}^{2N_t}-2c{\cal E}^{N_t}\cos(\phi_4 N_t)},
\nonumber
\ena
where $~c=-1~$ ($~c=+1~$) holds for antiperiodic (periodic) 
boundary conditions in $~x_4~$ and
$$
{\cal E}= 1 + \frac{{\cal M}^2+{\cal K}^2}{2(1+{\cal M})}
- \frac{\sqrt{{\cal M}^2+{\cal K}^2}\sqrt{({\cal M}+2)^2+{\cal K}^2}}
{2(1+{\cal M})} ~;
$$
$$
{\cal M}=m_0+\sum_{l=1}^{3}\left(1-\cos \phi_l\right), \qquad
{\cal K}=\sqrt{\ \sum_{l=1}^{3}\sin^2 \phi_l}, \qquad m_0>0.
$$

If we put all $\phi_{\mu}=0$, the Eqs. (\ref{6a},~\ref{6b})
reproduce the results for the free fermion correlator \cite{CaBa}.

In practical computations the fermion field $\psi_x$ is rescaled with a factor
\\  $1/(4+m_0)^{1/2}$ and the bare mass value $m_0$ is replaced
by the hopping-parameter $\kappa$ by
\begin{equation} \label{7}
\kappa=\frac{1}{2(4+m_0)}.
\end{equation}

\section {Lorentz Gauge Fixing}
\setcounter{equation}{0}

A lattice discretization of the Lorentz (or Landau) gauge
fixing condition can be written as follows

\eq  \label{9}
\sum_{\mu}\left(\sin\theta_{x,\mu}-\sin\theta_{x-\hat{\mu},\mu}\right)=0.
\en
In practice, instead of solving this local condition one maximizes
iteratively the gauge functional
\eq  \label{10}
F[\theta]=\frac{1}{4V}\sum_{x,\mu} \cos\theta_{x,\mu}=\mbox{Max.}
\en
with respect to (periodic) gauge transformations
\eq  \label{11}
\theta_{x,\mu} \longrightarrow \theta^{\alpha}_{x,\mu}=\alpha_x+
\theta_{x,\mu}-\alpha_{x+\hat{\mu}} \quad {\rm mod\ }2\pi, \qquad
\alpha_x \in (-\pi,\ \pi].
\en

In our simulations the maximization of the gauge functional (\ref{10})
has been continued until both the mean and the local maximal
absolute values of the l.h.s. in eq.(\ref{9}) became less than some
small given numbers (in our case, $10^{-6}$ and $10^{-5}$, respectively).
In order to accelerate the maximization one can apply overrelaxation
optimized with respect to some parameter $\omega$ \cite{MaOl2}.

We call the algorithm {\it standard Lorentz gauge fixing}, if it consists
only of local maximization and overrelaxation steps.
It is well-known
that this procedure normally gets stuck into local maxima of the gauge
functional. The solutions corresponding to different local maxima
are called {\it Gribov} or {\it gauge copies}.

It is a common believe (see also \cite{Zwan})
that the Gribov problem has to be solved
by searching for the {\it global maximum} of the gauge functional (\ref{10})
providing the {\it best} gauge copy (or copies, in case of degeneracy).
In \cite{BoMiMPPe} we have shown that in order to reach the global maximum
we have necessarily to remove both the DDS and the
ZMM from the gauge fields.

DDS are identified as follows. The plaquette angle
(\ref{1}) can be decomposed \cite{ToDe}
$~\theta_{x,\mu\nu}=\overline{\theta}_{x,\mu\nu}+2\pi n_{x,\mu\nu}~$,
where the gauge invariant $\overline{\theta}_{x,\mu\nu} \in (-\pi,\pi]$
can be interpreted as physical (electro-) magnetic flux
and the discrete gauge dependent contribution
$~2 \pi n_{x,\mu\nu}, ~~n_{x,\mu\nu}=0,\pm 1,\pm 2$ represents a Dirac
string passing through the given plaquette if $~n_{x,\mu\nu} \ne 0~$
(the Dirac plaquette). A set of Dirac plaquettes providing a world
sheet of a Dirac string on the dual lattice is called Dirac sheet.
Double Dirac sheets (DDS) consist of two sheets with opposite flux orientation
which cover the whole lattice and close by the periodic boundary conditions.
Thus, they can easily be identified by counting the total number of
Dirac pla\-quettes $N^{(\mu\nu)}_{DP}$ for every plane $(\mu;\nu)$.
The necessary condition for the appearance of a DDS is that at least
for one of the six planes $(\mu;\nu)$ holds
\eq  \label{12}
N^{(\mu\nu)}_{DP} \ge 2\frac{V}{N_{\mu}N_{\nu}}.
\en
DDS can be removed by periodic gauge transformations.
But -- as it was demonstrated in \cite{BoMiMPPa} -- the standard Lorentz gauge
fixing procedure usually does not succeed in doing this. DDS occur
quite independently of the lattice size and the chosen $\beta$.
As a consequence the numerical result for the non-zero momentum
transverse photon correlator significantly differs from the expected
zero-mass perturbative propagator \cite{BoMiMPPe,BoMiDD,Mitr1}.

The ZMM of the gauge field
\eq \label{13}
\phi_{\mu}=\frac{1}{V}\sum_{x}\theta_{x,\mu}
\en
do not contribute to the gauge field action (\ref{1}) either.
For gauge configurations representing a small fluctuation around
constant modes
$$
\theta_{x,\mu}=\phi_{\mu}+\delta \theta_{x,\mu}, \qquad \sum_{x}\delta%
\theta_{x,\mu}= 0, \qquad \left|\delta \theta_{x,\mu}\right| \ll 1,
$$
it is easy to see, that the maximum of the functional (\ref{10}) requires
$~\phi_{\mu}\equiv 0~$.
The latter condition can be achieved by non-periodic gauge transformations
\eq  \label{15}
\theta_{x,\mu} \longrightarrow \theta^{\ c}_{x,\mu}=c_{\mu}+\theta_{x,\mu}
\quad {\rm mod\ }2\pi, \qquad c_{\mu} \in (-\pi,\ \pi].
\en

Therefore, we realize the gauge fixing procedure as proposed in \cite{BoMiMPPe}.
The successive Lorentz gauge iteration steps are always
followed by non-periodic gauge transformations suppressing the ZMM.
At the end we check, whether the gauge field contains
yet DDS. The latter can be excluded simply by repeating the same
algorithm starting again with a random gauge transformation applied to
the same gauge field configuration.
We call the combined procedure {\it zero-momentum Lorentz gauge} (ZML gauge).
It provides with $99.99\%$ probability the global maxi\-mum of the gauge
functional. The photon propagator does perfectly agree with the expected
perturbative result throughout the Coulomb phase.

\section{Results}
\setcounter{equation}{0}

We consider qQED within the Coulomb phase at $~\beta~$ values between
2 and 10 for $~\kappa~$ values not too close to $~\kappa_c~$.
Monte Carlo simulations were carried out with a filter heat bath
method. In order to extract the pure ZMM effect, we first
apply the standard Lorentz gauge procedure modified by initial random gauges
in order to suppress DDS. Let us abbreviate the notation for this modified
Lorentz gauge procedure by LG. We compare the result with that for the ZML
gauge described above.

\begin{figure}[pt]
\begin{center}
\vskip -1.5truecm
\leavevmode
\hbox{
\epsfysize=14cm
\epsfxsize=14cm
\epsfbox{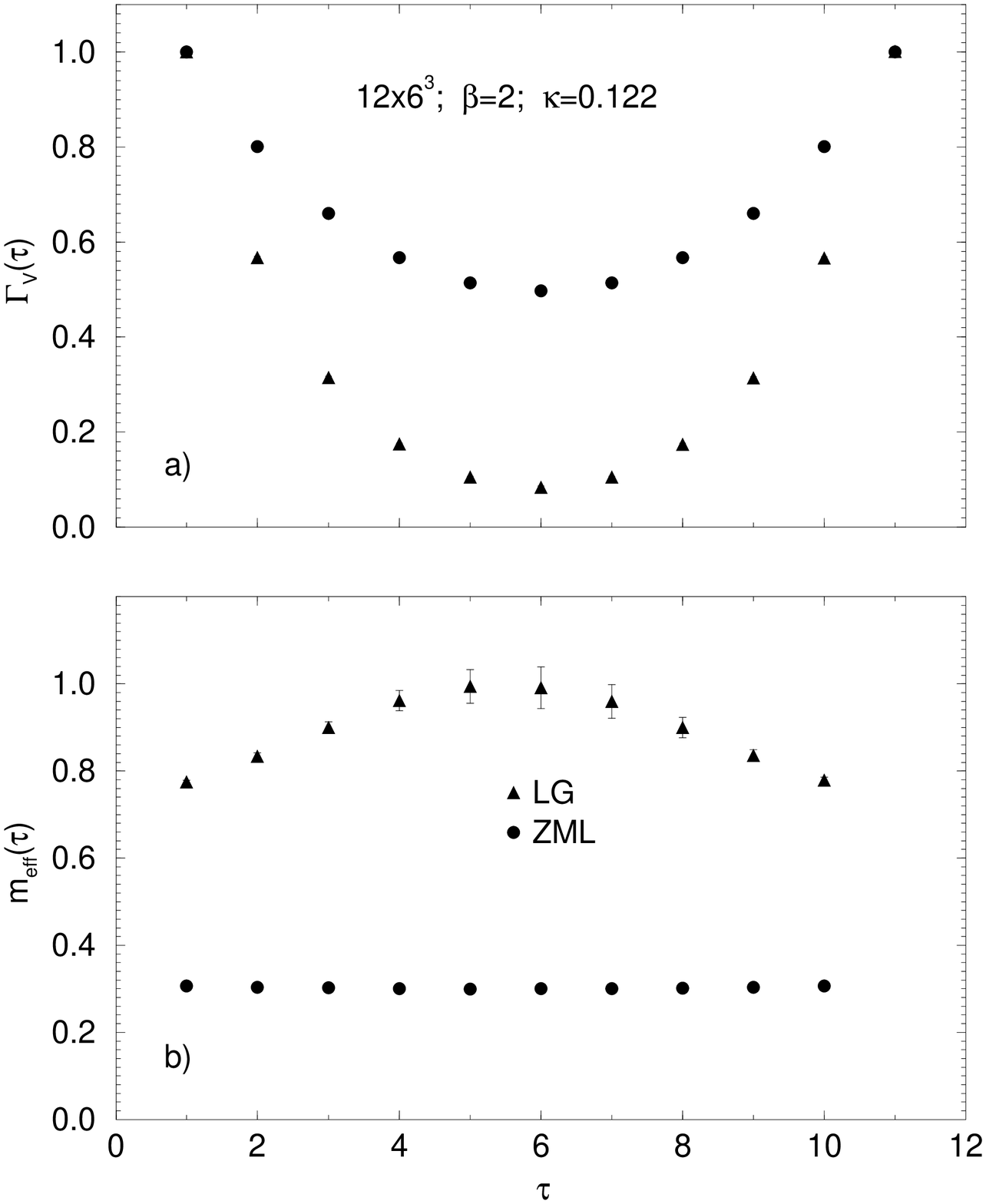}
}
\end{center}
\caption{The fermionic vector propagator ({\bf a}) and the effective
mass ({\bf b}) at $\beta=2.$ and $\kappa=0.122$ on a
$12\times 6^3$ lattice for LG and ZML gauges as explained in the text.
}
\label{fig:fc_meff_12x06_b02p00_k122_2gau}
\end{figure}

For both these gauges we have computed the averaged fermion correlator as
defined in Eqs. (\ref{4a},~\ref{4b}) and normalized to unity at $\tau=1$.
For inverting the Wilson-fermion matrix we
employed the conjugate gradient method and point-like sources.
In the upper part of Fig. \ref{fig:fc_meff_12x06_b02p00_k122_2gau} we have
plotted the vector part $~<\Gamma_V(\tau,\theta)>_{\theta}~$
for $~\beta=2$, $~\kappa=0.122~$
and lattice size $~12 \times 6^3~$. The situation seen is typical for
a wide range of parameter values within the Coulomb phase. Obviously,
there is a strong dependence of the fermion propagator on the gauge copies
differing by the different account of ZMM. If ZMM are present
the propagator decays much stronger than when they become suppressed.

The masses to be extracted seem to have different values. However,
let us try to extract the fermion mass as it is usually done with
an effective mass $~m_{eff}(\tau)~$ determined from the vector part
of the free fermion propagator (Eq. (\ref{6b}) with 
$~\phi_{\mu}=0~$).  I.e. we put

\eq  \label{16a}
\frac{\langle\Gamma_V(\tau+1;\theta)\rangle_{\theta}}
     {\langle\Gamma_V(\tau;  \theta)\rangle_{\theta}}
 =\frac{\cosh~[\ln (m_{eff}(\tau)+1) (N_t/2-\tau-1)]}
       {\cosh~[\ln (m_{eff}(\tau)+1) (N_t/2-\tau  )]}~.
\en

In the lower part of Fig. \ref{fig:fc_meff_12x06_b02p00_k122_2gau}
the corresponding numerical results for the effective masses are shown.
In the standard LG case no real plateau is visible, whereas
the ZML case provides a very stable one. Thus, the ZML gauge
yields a reliable mass estimate, whereas the standard case
fails. Naively, when only considering the LG method, one would be tempted
to relate a 'bad plateau' to finite-size effects and to believe that the
given LG effective mass result is already near to the real mass.
Such a point of view -- met in the literature -- obviously
fails. Taking now the ZML mass estimate as the reliable one
the LG estimate fails by a factor $~\sim 3$, in our case.

\begin{figure}[pt]
\begin{center}
\vskip -1.5truecm
\leavevmode
\hbox{
\epsfysize=14cm
\epsfxsize=14cm
\epsfbox{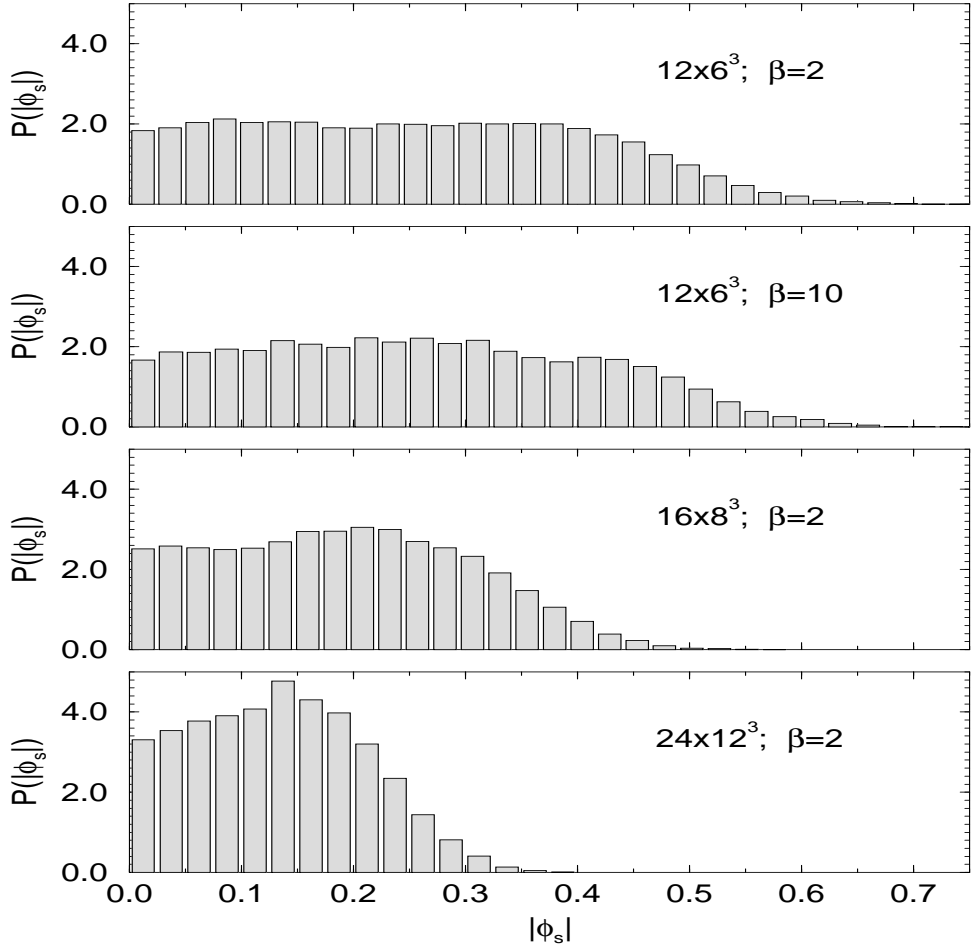}
}
\end{center}
\caption{Distributions of the spacelike zero--momentum mode
at different $\beta$--values and lattice sizes.
}
\label{fig:dist_2lat_2bet}
\end{figure}

One might ask, whether the troubles with the LG method disappear,
when we increase $~\beta~$ and/or the lattice size. In order to answer
this question, we first check, how the ZMM-distribution changes with
$~\beta~$ and with the lattice size. We have measured the distributions
of the moduli of the space- and time-like ZMM
$P(|\phi_s|)$ and $P(|\phi_4|)$, respectively,
for the LG case (with DDS suppressed).
In Fig. \ref{fig:dist_2lat_2bet} the corresponding space-like ZMM
distributions are drawn. The distributions turn out always
to be approximately step--like and  bound by
$~|\phi_{\mu}|~\leq~\phi_{max} \sim \pi/N_{\mu}$ with an
average value 

\eq
\langle |\phi_{\mu}|\rangle \sim \frac{\pi}{2N_{\mu}}~.
                \label{avzm}
\en

In order to estimate roughly the effect of the ZMM on the
fermion propagator for varying $\beta$ and lattice size we consider
the zero-momentum approximation as follows.
According to Eqs. (\ref{6a},~\ref{6b})
we compute the fermion propagator only within the constant background
modes extracted from the quantum gauge fields in the LG case
with the distribution $~P(\phi_{\mu})~$. Therefore, we compute

\eq
\frac{\langle\Gamma(\tau;\phi)\rangle_{\phi}}
     {\langle\Gamma(1;\phi)\rangle_{\phi}},
\qquad \langle\Gamma(\tau;\phi)\rangle_{\phi} =
\int [{\rm d} \phi] P(\phi) \Gamma(\tau;\phi) / \int [{\rm d} \phi] P(\phi),
                 \label{avcor}
\en

\begin{figure}[pt]
\begin{center}
\vskip -1.5truecm
\leavevmode
\hbox{
\epsfysize=14cm
\epsfxsize=14cm
\epsfbox{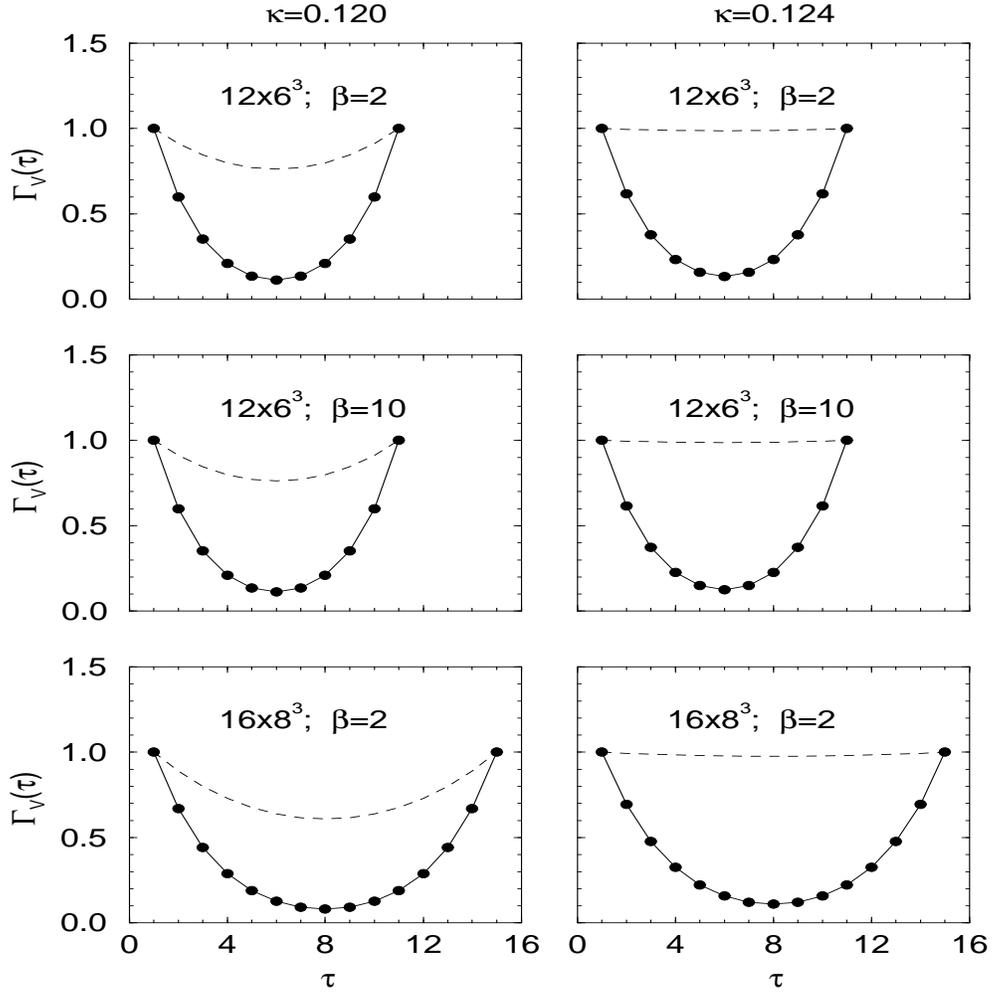}
}
\end{center}
\caption{Free fermionic vector propagator (dashed line)
and averaged constant-mode propagator (full line)
for two $\beta$--values and lattice sizes $12\times 6^3$,
$16\times 8^3$.
}
\label{fig:fc_tree_2lat_2bet}
\end{figure}

\noi where $\Gamma$ stands for the scalar or  vector part of the fermion
correlator (\ref{6a},~\ref{6b}).
The fermion mass $m_0$ is related to $\kappa$ according to Eq. (\ref{7}).
The results of this calculation for the vector part
of the fermion propagator in the LG case are presented
in Fig. \ref{fig:fc_tree_2lat_2bet} together with the corresponding
free (i.e. zero-background) propagator (dashed lines).
One can see that the effect of the ZMM does not 
weaken with increasing $\beta$ and lattice size, respectively.  
Given the average $~\langle |\phi_{\mu}|\rangle~$ as in Eq.(\ref{avzm})
one finds from Eqs.(\ref{6a},~\ref{6b},~\ref{avcor}) that the ZMM effect 
does not disappear even in the limit $~N_{\mu}\to\infty~$.

Preliminary Monte Carlo computations of the fermion propagator
within the full gauge field background confirm these observations.

\begin{figure}[pt]
\begin{center}
\vskip -1.5truecm
\leavevmode
\hbox{
\epsfysize=14cm
\epsfxsize=14cm
\epsfbox{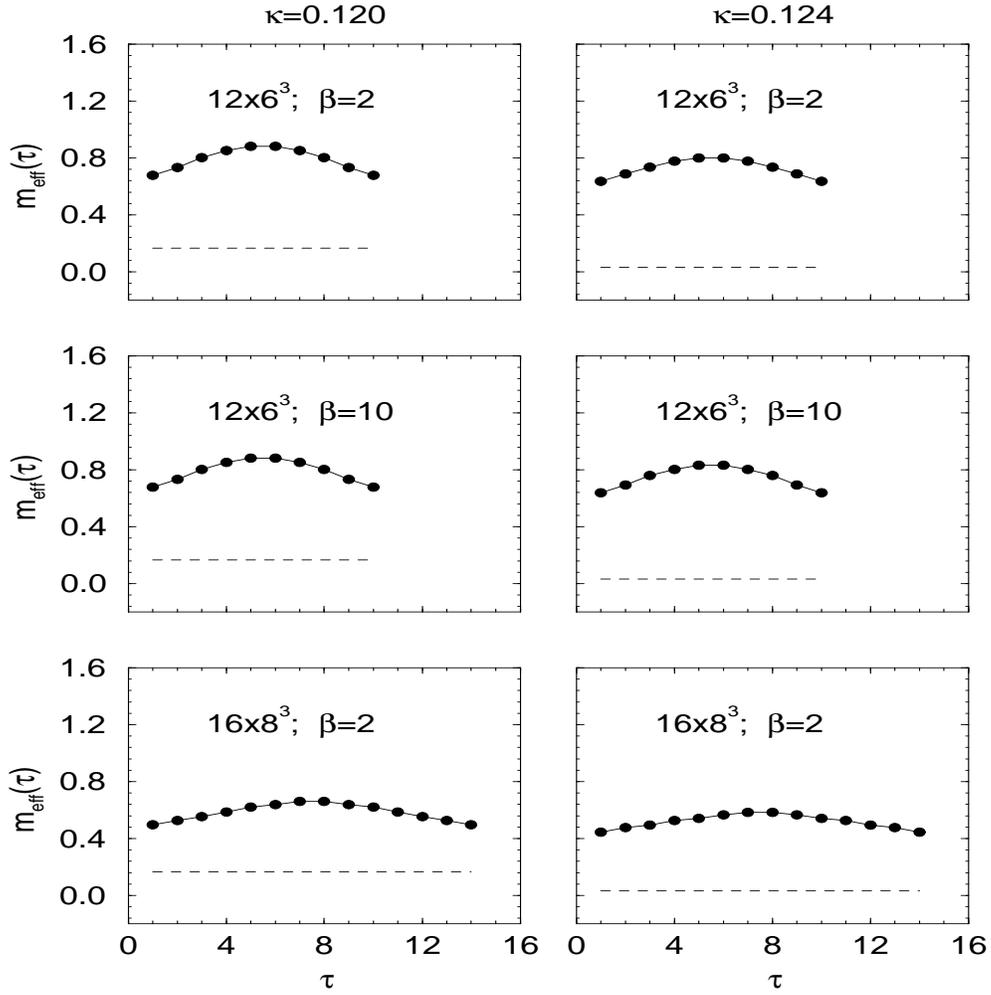}
}
\end{center}
\caption{Effective masses corresponding to the fermion propagator 
results shown in Fig. \ref{fig:fc_tree_2lat_2bet}.
}
\label{fig:meff_tree_2lat_2bet_2kap}
\end{figure}

We can take the approximation
described above in order to check,
how the corresponding effective fermion mass would behave.
This result is shown in Fig. \ref{fig:meff_tree_2lat_2bet_2kap}.
We clearly see, that for the LG case providing the ZMM
background field configurations 
we do not
find a plateau (full lines). The effective mass values strongly
differ from the real ones, i.e. $m_0$ of the free propagator
(dashed lines).

\begin{figure}[pt]
\begin{center}
\vskip -1.5truecm
\leavevmode
\hbox{
\epsfysize=14cm
\epsfxsize=14cm
\epsfbox{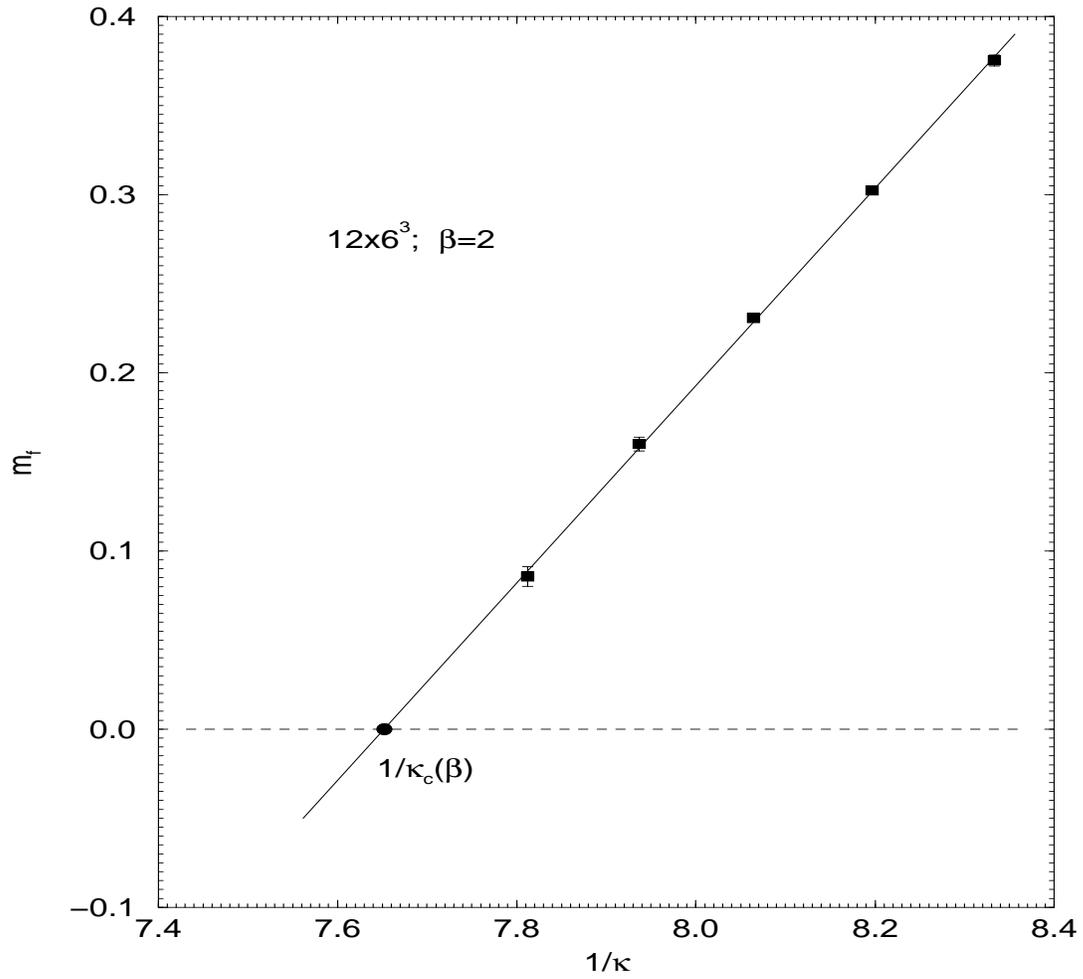}
}
\end{center}
\caption{Fermion mass as a function of inverse $\kappa$ obtained within
the ZML gauge for $\beta=2.0$ on a $12 \times 6^3$ lattice.
The solid line represents a linear fit providing 
$~\kappa_c(\beta) = 0.1307 \pm 0.0001~$. 
}
\label{fig:mf_fpr}
\end{figure}

Finally, in Fig. \ref{fig:mf_fpr} we present the fermion mass extracted
from the vector fermion propagator within the ZML gauge for
$\beta=2.0$ and various $\kappa$-values. 
We see a nice linear behaviour
from which by extrapolating to zero mass (solid line) we estimate 
the critical value $\kappa_c= 0.1307 \pm 0.0001$.  

\section{Conclusions}
\setcounter{equation}{0}

We have studied the special effect of the zero-momentum modes of the gauge
field on the gauge dependent fermion correlator.

We have convinced ourselves that the standard Lorentz gauge fixing
prescription to maximize the functional (\ref{10}) provides gauge
copies with ZMM (besides DDS). These modes disturb the
fermion correlator in comparison with  perturbation theory
and consequently spoil the (effective) mass estimate.
A Lorentz gauge employing non-periodic gauge transformations in order to
suppress the ZMM -- additionally to DDS --
(the ZML gauge) allows to reach the global maximum of the Lorentz gauge
functional. Furthermore, it  provides a reliable fermion mass determination,
at least, if $\kappa$ is chosen not too close to the chiral critical
line $\kappa_c(\beta)$. A
computation of the fermion propagator with constant background
gauge fields taken from the ZMM of the quantum fields
demonstrates the disturbing effect of these modes very clearly. Moreover,
it shows the effect to be independent of the bare coupling
and not to disappear for large volumes.

\bigskip
So far, we have studied the quenched approximation of U(1) lattice gauge theory.
The gauge action (\ref{1}) is invariant under non-periodic gauge
transformations (\ref{15}). Thus, we are allowed to use the ZML gauge
for evaluating gauge dependent objects.
Contrary to the gauge action, the fermionic part (\ref{2}) does depend
on the ZMM because of the (anti-) periodic boundary conditions.
In this case another solution of dealing with the Gribov problem
has to be searched for. One way, nevertheless, could be considering
standard LG and taking the constant background modes properly into account
in describing the perturbative finite-volume fermion propagator and
then identifying correspondingly the renormalized fermion mass
(see \cite{Hors}). This is under consideration now.

\section*{Acknowledgements}
The authors are grateful to A. Hoferichter for help and discussions.
Two of us (M.M.-P. and N.V.Z.) acknowledge discussions also with
R. Horsley and W. Kerler.
The work has been supported by the grant INTAS-96-370,
the RFRB grant 99-01-01230 and the JINR Dubna Heisenberg-Landau program.


%
%

\vfill


%
%

\vfill


%
%

\vfill


%
%

\vfill


%
%

\vfill

\end{document}